\documentclass[journal,a4paper]{IEEEtran}
\usepackage[final]{graphicx}
\usepackage{color}
\makeatletter
\def\E{\ensuremath{\mbox{\rm E}}}
\def\H{{\mathsf{H}}}
\def\T{{\mathsf{T}}}

\def\ul#1{\underline{#1}}
\def\NT{N_\mathrm{T}}
\def\NR{N_\mathrm{R}}
\def\ll{{[l]}}
\def\fl{{[\,\bar{l}\,]}}
\def\defeq{\stackrel{\mbox{\smash{\scriptsize def}}}{=}}
\def\diag{\textbf{diag}}

\def\Re{\rlap{$\mathcal{R}$}\hbox to .7em{\hss}\,}
\def\Im{\rlap{$\mathcal{I}$}\hbox to .7em{\hss}\,}
\def\ve#1{{\mathchoice{\mbox{\boldmath$\displaystyle #1$}}%
                      {\mbox{\boldmath$\textstyle #1$}}%
                      {\mbox{\boldmath$\scriptstyle #1$}}%
                      {\mbox{\boldmath$\scriptscriptstyle #1$}}}}

\def\intertext#1{\noalign{%
        \penalty\postdisplaypenalty\vskip\belowdisplayskip
        \vbox{\normalbaselines\noindent#1}%
        \penalty\predisplaypenalty\vskip\abovedisplayskip}}
\DeclareSymbolFont{AMSb}{U}{msb}{m}{n}
\DeclareFontFamily{U}{msb}{}%
\DeclareFontShape{U}{msb}{m}{n}{<-6>msbm5<6-8>msbm7<8->msbm10}{}%

\def\Z{\hbox{\usefont{U}{msb}{m}{n}\selectfont Z}}
\makeatother
\begin{document}
\title{Notes on Lattice-Reduction-Aided MMSE Equalization}
\author{Robert F.H. Fischer%,~\IEEEmembership{Senior Member,~IEEE}%
\thanks{R. Fischer is with
        Lehrstuhl f{\"u}r Informations\-{\"u}bertra\-gung,
        Friedrich--Alexander--Universit{\"a}t Erlangen--N{\"u}rnberg,
        Cauerstrasse 7/LIT, 91058 Erlangen, Germany, Email:
        \texttt{fischer@LNT.de}}
}
\markboth{ }{ }
\maketitle
\begin{abstract}
Over the last years, novel low-complexity approach\-es to the equalization of
MIMO channels have gained much attention. Thereby, methods based on lattice
basis reduction are of special interest, as they achieve the optimum diversity
order. In this paper, a tutorial overview on LRA equalization optimized
according to the MMSE criterion is given. It is proven that applying the
zero-forcing BLAST algorithm to a suitably augmented channel matrix---the
inverse of the square root of the correlation matrix of the data symbols times
the noise variance forms its lower part---results in the optimum solution.
This fact is already widely used but lacks a formal proof. It turns out that
it is more important to take the correlations of the data correctly into
account than what type of lattice reduction actually is used.
\end{abstract}
%
%%%%%%%%%%%%%%%%%%%%%%%%%%%%%%%%%%%%%%%%%%%%%%%%%%%%%%%%%%%%%%%%%%%%%%%%%%%%%%%
\section{Introduction}
\label{sec1}

\noindent
The joint reception of signals transmitted in parallel---either considering
multi-antenna systems or multi-user scenarios---will become even more important
over the next years. When designing transmission systems for such
\emph{mul\-tiple-in\-put/mul\-tiple-out\-put (MIMO) channels}, the interference
amoung the individual signals has to be dealt with by means of equalization.

During the last decade, numerous techniques known from
inter\-symbol-inter\-ference channels---e.g., linear equali\-zation,
de\-cision-feedback equalization (DFE, also known as \ul{s}ucces\-sive
\ul{i}n\-ter\-fer\-ence \ul{c}ancellation (SIC) and also used in the \ul{B}ell
\ul{La}bor\-a\-tories \ul{s}pace-\ul{t}ime (BLAST) approach),
maxi\-mum-likeli\-hood detection,
cf.\ \cite[Table~E.1]{Fischer:02}---have been transferred to the MIMO setting.
However, novel approaches based on lattice basis reduction, e.g.,
\cite{Yao:02,Windpassinger:03}, are of special interest. Using these
\emph{\ul{l}attice-\ul{r}eduction-\ul{a}ided (LRA) techniques},
low-com\-plexity equalization achieving the optimum diversity behavior
\cite{Taherzadeh:07} is enabled.

In this paper, a tutorial overview on LRA equalization optimized according
to the \emph{\ul{m}inimum \ul{m}ean-\ul{s}quared \ul{e}rror (MMSE)} criterion
is given. It is shown that it is more important to take the inherently
introduced correlations of the data symbols correctly into account, than which
lattice reduction approach actually is used. The main result of the paper is
to establish a connection of the V-BLAST algorithm to the successive MMSE
estimation of correlated data starting from basic principles. It is proven
that applying (zero-forcing) BLAST to a suitably augmented channel
matrix---having the inverse of the square root of the correlation matrix of
the data symbols times the noise variance as its lower part---indeed results
in the optimum solution. To the best knowledge of the authors, a formal proof
for this fact yet has not being presented in literature. However, it has been
used widely without taking care of its validity.

The paper is organized as follows: in Sec.~\ref{sec2} the channel model
is introduced and conventional equalization techniques are briefly reviewed.
Lattice-reduction-aided equalization is addressed in Sec.~\ref{sec3} and
its MMSE DFE version is analyzed in detail in Sec.~\ref{sec4}. Concluding
remarks follow in Sec.~\ref{sec5} and in the Appendix fundamentals on
estimation are compiled.

%
%
%

%%%%%%%%%%%%%%%%%%%%%%%%%%%%%%%%%%%%%%%%%%%%%%%%%%%%%%%%%%%%%%%%%%%%%%%%%%%%%%%
\section{Channel Model and Equalization}
\label{sec2}

\noindent
We consider uncoded multiple-antenna transmission over flat-fading channels
where joint equalization at the receiver side is possible. The input/output
relation is given by the usual equation%
%%%%
\footnote{Notation: $\ve{A}^\T$: transpose of matrix $\ve{A}$;
        $\ve{A}^\H$: Hermitian (i.e., conjugate) transpose;
        $\ve{A}^{-\H}$: inverse of the Hermitian transpose of a
        square matrix $\ve{A}$; $\ve{I}$: identity matrix;
	matrices are denoted by uppercase letters, vectors by lower case
	letters.
	$\E\{ \cdot \}$: expectation.
}
%%%%
\begin{equation}                                                 \label{eq_2_1}
\ve{y} = \ve{H}\ve{a} + \ve{n} \;.
\end{equation}
This model has either to be understood in complex base-band notation or as
its real-valued model with doubled dimensionality \cite{Hassibi:00}.
As all subsequent discussions can either be applied to the complex or the real
model, we do not distinguish both approaches in the sequel. In each case,
the channel matrix is expected to be of dimension $\NR \times \NT$.
The differences between both views are discussed if required. For successive
schemes, due to the larger degree of freedom, usually the real-valued model
has some advantages \cite{Fischer:03reell}.

Each component $a_{\mu}$ of $\ve{a}$ is independently drawn from a
zero-mean one-dimensional $M$-ary ASK constellation $\mathcal{A} = \{ \pm 1/2,
	\pm 3/2, \ldots, \pm (M-1)/2 \}$ or an $M^2$-ary QAM constellation,
with an $M$-ary ASK per quadrature component. The correlation matrix of the
data vector hence reads $\ve{\Phi}_{aa} \defeq \E\{ \ve{a}^{}\ve{a}^\H \}
= \sigma_a^2\ve{I}$, with variance $\sigma_a^2 \defeq \E\{ |a_{\mu}|^2 \}$.
The noise is assumed to be spatially white with variance $\sigma_n^2$ per
component, i.e., $\ve{\Phi}_{nn} \defeq \E\{ \ve{n}^{}\ve{n}^\H \}
= \sigma_n^2\ve{I}$.

\subsection{Linear Equalization}

The interference between the parallel data streams can be eliminated by
means of equalization, i.e., via $\ve{r} = \ve{H}_\mathrm{R}\ve{y}$ a decision
vector is generated. Having $\ve{r}$, individual threshold decision can be
performed.

Using \emph{\ul{l}inear \ul{e}qualization (LE)}, optimized according to the
\emph{\ul{z}ero-\ul{f}orcing (ZF)} criterion, the receive matrix reads
\begin{equation}                                                \label{eq_2_10}
\ve{H}_\mathrm{R}^{(\mathsf{LE,ZF})}
		= \left(\ve{H}^\H\ve{H}\right)^{-1} \ve{H}^\H \;,
\end{equation}
i.e., the receive matrix is given by the Moore-Penrose left pseudo inverse
of $\ve{H}$. Already in \cite{Hassibi:00} it has been observed that the
minimum mean-squared error solution is obtained by using the augmented matrix
($\zeta \defeq \frac{\sigma_n^2}{\sigma_a^2}$ is the inverse signal-to-noise
ratio)
\begin{equation}                                                \label{eq_2_11}
\ve{\bar{H}} = \left[ \matrix{ \ve{H} \cr \sqrt{\zeta}\ve{I} } \right]_
				{(\NR+\NT)\times \NT}
	\qquad
	\ve{\bar{y}} = \left[ \matrix{ \ve{y} \cr \ve{0} } \right]_%
			       {(\NR+\NT)}
\end{equation}
in the ZF solution and feeding $\ve{\bar{y}}$ into the resulting receive
matrix rather than $\ve{y}$. Subsequently, all quantities corresponding to
the augmented channel model are marked by a horizontal bar.

\subsection{Decision-Feedback Equalization}

Some gains over linear equalization can be achieved by using sorted
decision-feedback equalization, also known as BLAST or SIC. The required
matrices for ZF DFE are obtained by performing a sorted QR-type decomposition
such that
\begin{equation}                                                \label{eq_2_20}
\ve{H}\ve{P} = \ve{Q}\ve{L} \;,
\end{equation}
where $\ve{P}$ is a permutation matrix (a single one in each row and column),
$\ve{Q}$ is unitary and $\ve{L}$ is lower triangular. From these quantities,
the feedforward matrix $\ve{F}$ and the lower triangular, unit main diagonal
feedback matrix $\ve{B}$ are calculated as%
%%%%
\footnote{$\diag(\cdot)$ denotes a diagonal matrix with elements taken from
	the main diagonal of the indicated matrix.
}
%%%%
$\ve{F} \defeq \diag(\ve{L})^{-1}\ve{Q}^\H$ and
$\ve{B} \defeq \diag(\ve{L})^{-1}\ve{L}$, respectively.

Again, the MMSE solution is obtained by plugging the augmented channel matrix
into (\ref{eq_2_20}), cf.\ \cite{Windpassinger:04}.

%
%
%

%%%%%%%%%%%%%%%%%%%%%%%%%%%%%%%%%%%%%%%%%%%%%%%%%%%%%%%%%%%%%%%%%%%%%%%%%%%%%%%
\section{Lattice-Reduction-Aided Equalization}
\label{sec3}

\noindent
Unfortunately, using linear equalization or DFE, only a diversity order of
$\NR-\NT+1$ (for the complex-valued model) is possible. Lattice-reduction-aided
equalization schemes, e.g., \cite{Yao:02,Windpassinger:03}, have proven to
require only low complexity, nevertheless being able to achieve the full
diversity order $\NR$ of the MIMO channel \cite{Taherzadeh:07}. The idea is
to choose a ``more suited'' representation of the lattice spanned by the
columns of the channel matrix $\ve{H}$; equalization is done with respect to
the new basis, which is desired to be close to orthogonal. At the very end,
the change of basis is reversed.

\subsection{Lattice-Reduction-Aided Linear Equalization}

For performing LRA equalization, in the first step lattice basis reduction,
e.g., by using the LLL algorithm \cite{Lenstra:82} (or some complex-valued
version thereof, e.g., \cite{Gan:09}), is performed to obtain
\begin{equation}                                                 \label{eq_3_1}
\ve{H} = \ve{C}\,\ve{Z} \;,
\end{equation}
where $\ve{Z} \in \Z^{\NT\times\NT}$ is an integer unimodular matrix, i.e.,
has only integer coefficients%
%%%%
\footnote{In case of complex signals, the set of integers $\Z$ has to be
	replaced by the set $\Z + \mathrm{j}\Z$ of Gaussian integers.
}
%%%%
and $|\mathrm{det}(\ve{Z})| = 1$. The reduced channel matrix $\ve{C}$ is usually
required to have columns close to orthogonal and of small norms (depending
on the definition of ``reduced''). Using (\ref{eq_3_1}), the receive signal is
given by $\ve{y} = \ve{C}\ve{Z}\ve{a} + \ve{n}$.

In the second step, only $\ve{C}$ is treated and the signal
$\ve{z} \defeq \ve{Z}\ve{a}$, which is taken from a translate of the integer
lattice ($\ve{Z}\Z^{\NT} = \Z^{\NT}$) and hence can be obtained by individual
threshold decision per component, is to be estimated. This transformed data
vector has zero mean, $\ve{\mu}_z = \E\{ \ve{z} \} = \E\{ \ve{Z}\ve{a} \}
= \ve{Z}\E\{ \ve{a} \} = \ve{Z}\ve{0} = \ve{0}$, but is correlated with
covariance matrix
\begin{equation}                                                 \label{eq_3_2}
\ve{\Phi}_{zz} = \E\{ \ve{z}\ve{z}^\H \}
	= \E\{ \ve{Z}\ve{a}\ve{a}^\H\ve{Z}^\H \}
	= \sigma_a^2 \ve{Z}\ve{Z}^\H \;.
\end{equation}
Third, the change of basis is reversed via $\ve{Z}^{-1}$.

\subsubsection{LRA ZF Linear Equalization}

Applying LRA ZF linear equalization the correlations are ignored and the
receive matrix is simply the left pseudo inverse of the reduced channel matrix
\begin{equation}                                                 \label{eq_3_3}
\ve{H}_\mathrm{R}^{(\mathsf{LRA,LE,ZF})} =
	\left( \ve{C}^\H\ve{C} \right)^{-1} \ve{C}^\H \;.
\end{equation}

\subsubsection{LRA MMSE Linear Equalization}

As in the conventional case, the MMSE solution may be obtained by applying
all operations to the augmented channel model, cf.\ \cite{Wuebben:04}.
Hence, in the first step $\ve{\bar{H}}$ is fed into the lattice basis
reduction, resulting in (note: $\ve{Z}$ usually differs from the ZF case)
\begin{equation}                                                \label{eq_3_10}
\ve{\bar{H}} = \ve{\bar{C}}\,\ve{Z} \;.
\end{equation}
Using the definition of $\ve{\bar{H}}$, the reduced augmented matrix can be
written as
\begin{equation}                                                \label{eq_3_11}
\ve{\bar{C}}
	= \left[ \matrix{ \ve{H} \cr \sqrt{\zeta}\ve{I} } \right]
			\ve{Z}^{-1}
	= \left[ \matrix{ \ve{H}\ve{Z}^{-1} \cr
			\sqrt{\zeta}\ve{Z}^{-1} } \right]
	\defeq \left[ \matrix{ \ve{C} \cr \ve{A} } \right] \;,
\end{equation}
with the obvious definitions of $\ve{C}$ and $\ve{A}$.
The receive matrix (with respect to $\ve{\bar{y}}$) is then given by
\begin{eqnarray}                                                \label{eq_3_12}
\ve{\bar{H}}_\mathrm{R}^{(\mathsf{LRA,LE,MMSE})}
	&=& (\ve{\bar{C}}^\H\ve{\bar{C}}^{})^{-1} \ve{\bar{C}}^\H
\\
&& \hskip-15mm = \left( \ve{C}^\H\ve{C}^{} + 
		\zeta \ve{Z}^{-\H}\ve{Z}^{-1} \right)^{-1}
	\left[ \matrix{ \ve{C}^\H \; \sqrt{\zeta}\ve{Z}^{-\H} } \right] \;,
\nonumber
\end{eqnarray}
or with respect to $\ve{y}$, when deleting the last $\NT$ columns
\begin{eqnarray}                                                \label{eq_3_13}
\ve{H}_\mathrm{R}^{(\mathsf{LRA,LE,MMSE})} &=&
	\left( \ve{C}^\H\ve{C}^{} + \zeta \ve{Z}^{-\H}\ve{Z}^{-1} \right)^{-1}
		\ve{C}^\H
\\
&=& \ve{Z} \left( \ve{H}^\H\ve{H}^{} + \zeta\ve{I} \right)^{-1} \ve{H}^\H \;.
\end{eqnarray}

This receive matrix takes the correlations of the data perfectly into account.
To see this, note that from the basic literature on estimation, e.g.,
\cite[Theorem 2.6.1]{Sayed:03}, the optimum MMSE linear estimator is given by
\begin{equation}                                               \label{eq_3_13a}
\left( \ve{C}^\H \ve{\Phi}_{nn}^{-1} \ve{C}^{} + \ve{\Phi}_{zz}^{-1} \right)^{-1}
			 \ve{C}^\H\ve{\Phi}_{nn}^{-1} \;,
\end{equation}
which, since white channel noise was assumed and $\ve{\Phi}_{zz} =
	\sigma_a^2 \ve{Z}\ve{Z}^\H$, exactly gives the receive matrix
(\ref{eq_3_13}). The covariance matrix of the resulting minimum mean-squared
error $\ve{e}$ is given by
\begin{eqnarray}                                                \label{eq_3_14}
\ve{\Phi}_{ee} &=& \left( \ve{C}^\H \ve{\Phi}_{nn}^{-1}
		\ve{C}^{} + \ve{\Phi}_{zz}^{-1} \right)^{-1}
\nonumber\\
	&=& \sigma_n^2\left( \ve{C}^\H \ve{C}^{} +
		\zeta \ve{Z}^{-\H}\ve{Z}^{-1} \right)^{-1} \;.
\end{eqnarray}

%Offset compensation and decision are carried out as in the ZF case.
%Noteworthy, the signal to be estimated is in principle unbounded and may be
%taken from the entire integer grid. Hence, no bias compensation, opposed to
%the finite case, is required or possible.

%
%
%
\subsection{Lattice-Reduction-Aided ZF DFE}

\begin{figure}[htb]
\centerline{\includegraphics{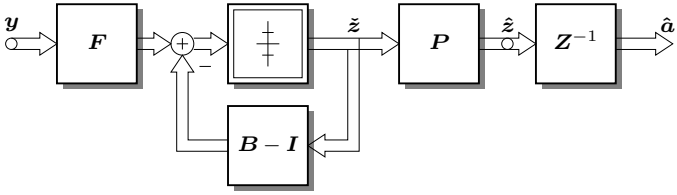}}
\caption{\label{fig_3_3}
	Lattice-reduction-aided DFE.
}
\end{figure}

In order to enhance performance, linear equalization can be replaced
by DFE, resulting in \emph{lattice-reduction-aided DFE}, cf.\ Fig~\ref{fig_3_3}.
%
%%%%
%%%%

As in the classical case, for performing DFE, the (sorted) QR-type
factorization of the respective channel matrix is required. For LRA ZF DFE,
the factorization has the form
\begin{equation}                                                \label{eq_3_20}
\ve{C}\ve{P} = \ve{Q}\ve{L} \;.
\end{equation}
Feedforward and feedback matrices $\ve{F}$ and $\ve{B}$ are calculated as
explained above. In the feedback loop, the components of $\ve{z}$ are detected
in an optimized order described by the permutation matrix $\ve{P}$. After
reestablishing the original ordering, an estimate of the original data vector
$\ve{a}$ is generated via the inverse of the integer unimodular matrix
$\ve{Z}$.

%
%
%

%%%%%%%%%%%%%%%%%%%%%%%%%%%%%%%%%%%%%%%%%%%%%%%%%%%%%%%%%%%%%%%%%%%%%%%%%%%%%%%
\section{Lattice-Reduction-Aided MMSE DFE}
\label{sec4}

\noindent
The optimization of the LRA DFE according to the MMSE criterion is not as
straightforward as in the ZF case. This is due to the correlation of the data
symbols $z_k$ to be estimated in an optimum succession within the DFE loop.
Up to now, in the literature this fact has not been treated in detail; usually
simply the ZF solution with respect to the augmented matrix has been used, e.g.,
\cite{Wuebben:04,Murugan:06}. We first review the straightforward application
of the BLAST algorithm \cite{Golden:99} to the augmented channel model and
then compare these results to those obtained from the theory of optimum
estimation of correlated Gaussian random variables.

\subsection{Lattice Reduction}

As in the LRA MMSE linear case, we stick to the augmented channel model
$\ve{\bar{H}}$ and consider the lattice reduction according to (\ref{eq_3_10})
and (\ref{eq_3_11}). Assume for simplicity of notation, that the columns of
$\ve{\bar{C}}$ are sorted according to the optimum decision
order, i.e., we replace $\ve{C}$ implicitly by $\ve{C}\ve{P}$, thereby
anticipating the permutation matrix $\ve{P}$ to be determined during the
calculation of the required matrices. Thereby, the optimization criterion
is---as proposed in the V-BLAST system---the noise enhancement encountered
in the feedforward processing. For the MMSE solution this criterion is
identical to looking at the minimum main diagonal element of the error
covariance matrix.

%Decisions within the feedback loop are always taken in natural succession. 

%
%
%

\subsection{V-BLAST Algorithm}

We first simply perform the (MMSE) V-BLAST algorithm with respect to the
augmented channel matrix $\ve{\bar{C}}$. Assuming that $l$
($l=0,\,\ldots,\,\NT-1$) symbols are already known, the BLAST approach is to
simply delete the $l$ first columns (due to the assumed sorting) of
$\ve{\bar{C}}$ and proceed with the residual%
%%%%
\footnote{Given a matrix $\ve{M}$, let $\ve{M}_\ll$ denote the matrix
	obtained from $\ve{M}$ by deleting the first $l$ columns,
	and $\ve{M}_\fl$ denote the matrix composed of the first $l$ columns
	of $\ve{M}$, i.e., $\ve{M} = [\ve{M}_\fl\,\ve{M}_\ll]$.
	Please distinguish that from the notation $\cdot^{(l)}$, which indicates
	a quantity present in step $l$ (counting from zeros to $\NT-1$) of the
	algorithm.
}
%%%%
augmented channel matrix $\ve{\bar{C}}_\ll$.

\subsubsection{Feedforward Matrix}

Having deleted the first $l$ columns, the potential feedforward matrix
(with respect to the augmented channel model) for estimating the remaining
$\NT-l$ symbols reads
\begin{eqnarray}						\label{eq_4_10}
\ve{\bar{F}}^{(l)} &=& 
	\left[ \matrix{ \ve{\bar{f}}_1^{(l)} \cr \vdots \cr
				\ve{\bar{f}}_{\NT-l}^{(l)} }
									\right]
	\;=\; \left( \ve{\bar{C}}_\ll^\H \ve{\bar{C}}_\ll^{} \right)^{-1}
			\ve{\bar{C}}_\ll^\H
\nonumber\\
&=& \left( \ve{C}_\ll^\H \ve{C}_\ll^{} + \ve{A}_\ll^\H\ve{A}_\ll^{} \right)^{-1}
	\left[ \ve{C}_\ll^\H \, \ve{A}_\ll^\H \right] \;.
\end{eqnarray}
In each step the row $\ve{\bar{f}}_k^{(l)}$, corresponding to the symbol $z_k$
which can be detected most reliably, is appended to the entire feedforward
matrix $\ve{\bar{F}}$. The feedforward matrix for the non-augmented, original
channel is obtained from $\ve{\bar{F}}$ by deleting the last $\NT$ columns.

\subsubsection{Optimum Sorting}

In the BLAST algorithm, usually the norms of the row of the feedforward matrix
are considered as sorting criterion \cite{Golden:99}. These are proportional
to the noise enhancement and hence determine the error rate.
Using (\ref{eq_4_10}), these row norms are given by the diagonal elements of
\begin{eqnarray}						\label{eq_4_12}
\ve{\bar{F}}^{(l)}(\ve{\bar{F}}^{(l)})^\H &=&
	\left( \ve{\bar{C}}_\ll^\H \ve{\bar{C}}_\ll^{}\right)^{-1}
			\ve{\bar{C}}_\ll^\H 
		\left( \ve{\bar{C}}_\ll^\H \ve{\bar{C}}_\ll^{} \right)^{-\H}
\nonumber\\
&=& \left( \ve{\bar{C}}_\ll^\H \ve{\bar{C}}_\ll^{} \right)^{-1}
\nonumber\\
%&=& \left( \left[ (\ve{H}_\mathrm{red}^{[l]})^\H \, (\ve{A}^{[l]})^\H \right]
%		\left[ \ve{H}_\mathrm{red}^{[l]} \atop \ve{A}^{[l]} \right]
%			\right)^{-1}
%\nonumber\\
&=& \left( \ve{C}_\ll^\H \ve{C}_\ll^{} + \ve{A}_\ll^\H\ve{A}_\ll^{} \right)^{-1}
	\;.
\end{eqnarray}

If $\ve{\bar{C}}$ has already been sorted optimally, the upper left diagonal
element will be the smallest. Otherwise, the first row of $\ve{\bar{F}}$ and
that with the smallest norm are exchanged; this exchange is also recorded in the
permutation matrix $\ve{P}$. After $\NT$ iterations the entire feedforward
matrix $\ve{\bar{F}}$ and the optimum processing order, represented by the
permutation matrix $\ve{P}$ are known.

\subsubsection{Feedback Matrix}

Knowing $\ve{\bar{F}}$ and $\ve{P}$, the feedback matrix $\ve{B}$ can be
calculated. It is well-known \cite{Ginis:02} that the approaches of
a) canceling before applying the feedforward matrix (as usually proposed in
the BLAST context) and
b) canceling at the output of the feedforward matrix (as is preferred in the
DFE context) are equivalent. Here, we consider the latter strategy, cf.\
also Fig.~\ref{fig_3_3}.

Since it is optimum to cancel all known interference, the feedback matrix
calculates to
\begin{equation}						\label{eq_4_13}
\ve{B} = \left[ \matrix{ \ve{b}_1 \cr \vdots \cr \ve{b}_{\NT} } \right]
	= \ve{\bar{F}} \ve{\bar{C}} \ve{P}
	= \ve{\bar{F}} \left[ \matrix{\ve{C} \cr \ve{A}} \right] \ve{P} \;.
\end{equation}
As in each step $\ve{\bar{F}}_\ll^{}\ve{\bar{C}}_\ll^{} = \ve{I}$ holds, i.e.,
the remaining symbols are equalized and the already canceled are ignored, it is
easy to see that $\ve{B}$ is a lower triangular matrix with unit main diagonal.
Moreover, by construction, the rows of $\ve{\bar{F}}$ are orthogonal; via a
diagonal gain matrix $\ve{G}$ we can write
$\ve{\bar{F}} = \ve{G}\ve{\bar{Q}}^\H$, where $\ve{\bar{Q}}$ is an
$(\NR+\NT)\times \NT$ matrix with orthonormal columns. In summary, using the
lower triangular matrix $\ve{L} \defeq \ve{G}^{-1}\ve{B}$ (\ref{eq_4_13}) can
be written in the form
\begin{equation}						\label{eq_4_14}
\ve{\bar{C}} \ve{P} = \ve{\bar{Q}}\ve{L} \;,
\end{equation}
i.e., applying the BLAST algorithm a sorted QR-type (QL) factorization of the
reduced augmented channel matrix $\ve{\bar{C}}$ is inherently performed.

In more detail, the $(l+1)$th row of the feedback filter is given by
\begin{equation}						\label{eq_4_15}
\ve{b}_{l+1} = \ve{\bar{f}}_1^{(l)}\ve{\bar{C}}\ve{P} \;,
\end{equation}
which, using (\ref{eq_4_10}), is the first row of the matrix
\begin{eqnarray}						\label{eq_4_24}
\ve{\bar{M}}^{(l)} &=& \left( \ve{C}_\ll^\H \ve{C}_\ll^{}
		+ \ve{A}_\ll^\H\ve{A}_\ll^{} \right)^{-1} \!\!
	\left[ \ve{C}_\ll^\H \, \ve{A}_\ll^\H \right]\!\!
	\left[ \matrix{\ve{C} \cr \ve{A}} \right] \!\ve{P} \;.
\end{eqnarray}
Writing $\ve{\bar{C}}\ve{P} = \left[ {\ve{C}_1 \atop \ve{A}_1}
	{\ve{C}_2 \atop \ve{A}_2} \right]$, with $\ve{C}_2 = \ve{C}_\ll^{}$ and
$\ve{A}_2 = \ve{A}_\ll^{}$, we can write
\begin{eqnarray}						\label{eq_4_25}
\ve{\bar{M}}^{(l)}
	&=& \left( \ve{C}_2^\H \ve{C}_2^{} + \ve{A}_2^\H\ve{A}_2^{} \right)^{-1}
		\!\! \left[ \ve{C}_2^\H \, \ve{A}_2^\H \right]
			\left[ \matrix{ \ve{C}_1 & \!\!\!\ve{C}_2 \cr
				\ve{A}_1 & \!\!\!\ve{A}_2 } \right] 
\\
&& \hskip-10mm = \left[
	\left( \ve{C}_2^\H \ve{C}_2^{} + \ve{A}_2^\H\ve{A}_2^{} \right)^{-1}
	\!\! \left( \ve{C}_2^\H \ve{C}_1^{} + \ve{A}_2^\H\ve{A}_1^{} \right)
	\;\mid\; \ve{I} \, \right] \;.
\nonumber
\end{eqnarray}

\subsection{Optimum Estimation of Correlated Data}

We now turn to the situation of deriving the required matrices directly from
the theory of minimum mean-squared estimation and the properties of correlated
random vectors when parts of the variables are already known.
Looking at the optimum linear estimator (\ref{eq_a_1}), summarized in the
Appendix, feedforward and feedback matrices can immediately be given by
identifying the respective quantities suitably.

However, from (\ref{eq_a_1}) it can be deduced that the optimal processing
depends on the mean and covariance matrix of the vector of not yet detected
symbols. These quantities, however, depend on the previous decisions when
performing DFE. In turn, optimum filtering and the optimum processing order
potentially may depend on the actual decisions made so far within the DFE.
In the  following we show, that this is actually not the case. All required
matrices can be calculated in advance and the influence of previous decisions
is taken into account via the feedback matrix in an optimum way.

\subsubsection{Feedforward Matrix}

Again assume that the first $l$ symbols $z_k$ (contained in the vector
$\ve{z}_1$) have already been detected. We can partition the vector $\ve{z}$,
mean vector and correlation matrix of this vector in the form
\begin{equation}						\label{eq_4_40}
\ve{z} = \left[ \matrix{ \ve{z}_1 \cr \ve{z}_2 } \right]
	,\;
\ve{\mu}_z = \left[ \matrix{ \ve{\mu}_{1} \cr \ve{\mu}_{2} } \right]
	,\;
\ve{\Phi}_{zz} = \left[ \matrix{ \ve{\Phi}_{11} & \ve{\Phi}_{12} \cr
			\ve{\Phi}_{21} & \ve{\Phi}_{22} } \right] \;.
\end{equation}
Under white noise, the feedforward matrix---filtering the receive vector
$\ve{y}$ (non-augmented model) for obtaining estimates of the remaining
$\NT-l$ symbols---is given by (cf.\ (\ref{eq_a_1}))
\begin{eqnarray}						\label{eq_4_41}
\ve{F}^{(l)} &=&  
	\left[ \matrix{ \ve{f}_1^{(l)} \cr \vdots \cr \ve{f}_{\NT-l}^{(l)} }
									\right]
	= \Big( \ve{C}_\ll^\H \ve{C}_\ll^{} \textstyle\frac{1}{\sigma_n^2}
		+ \ve{\Phi}_{22 \mid \ve{z}_1}^{-1} \Big)^{-1} 
		\ve{C}_\ll^\H \textstyle\frac{1}{\sigma_n^2}
\nonumber\\
&=& \Big( \ve{C}_\ll^\H \ve{C}_\ll^{} + \sigma_n^2\ve{\Phi}_{22 \mid \ve{z}_1}^{-1}
			\Big)^{-1} \ve{C}_\ll^\H \;.
\end{eqnarray}

The conditioned covariance matrix $\ve{\Phi}_{22 \mid \ve{z}_1}^{-1}$ can be
written as follows. Since from (\ref{eq_3_11}) $\sqrt{\zeta}\ve{Z}^{-1} = \ve{A}
	\defeq [\ve{A}_1\,\ve{A}_2]$, we have on the one hand
\begin{eqnarray}						\label{eq_4_42}
\ve{\Phi}_{zz}^{-1} &=& \frac{1}{\sigma_a^2}\ve{Z}^{-\H}\ve{Z}^{-1}
	\;=\; \frac{1}{\sigma_n^2} \ve{A}^\H\ve{A}
\nonumber\\
&=& \frac{1}{\sigma_n^2} \left[ \matrix{
		\ve{A}_1^\H\ve{A}_1^{} & \ve{A}_1^\H\ve{A}_2^{} \cr
		\ve{A}_2^\H\ve{A}_1^{} & \ve{A}_2^\H\ve{A}_2^{} } \right] \;.
\end{eqnarray}
On the other hand, with the partitioning (\ref{eq_4_40}) and
using \cite[Page~472, Eq.~(7.7.5)]{Horn:85}, we can write (elements marked
by $*$ are irrelevant)
\begin{eqnarray}						\label{eq_4_43}
\ve{\Phi}_{zz}^{-1} &=& \left[ \matrix{ * & * \cr * &
		\left(\ve{\Phi}_{22}^{} - \ve{\Phi}_{21}^{}
			\ve{\Phi}_{11}^{-1}\ve{\Phi}_{12}^{}\right)^{-1} }
		\right] \;.
\end{eqnarray}
A comparison of (\ref{eq_4_42}) and (\ref{eq_4_43}) reveals that for all $l$,
we have
\begin{equation}						\label{eq_4_44}
\sigma_n^2\left(\ve{\Phi}_{22}^{} - \ve{\Phi}_{21}^{}
			\ve{\Phi}_{11}^{-1}\ve{\Phi}_{12}^{}\right)^{-1} \!\!
	= \ve{A}_2^\H\ve{A}_2^{}
	= \ve{A}_\ll^\H\!\ve{A}_\ll^{} .
\end{equation}
Hence, the optimum feedforward matrix calculates to
\begin{equation}						\label{eq_4_45}
\ve{F}^{(l)} =
	\left( \ve{C}_\ll^\H \ve{C}_\ll^{} + \ve{A}_\ll^\H\ve{A}_\ll^{}
		\right)^{-1} \ve{C}_\ll^\H \;.
\end{equation}

\subsubsection{Optimum Sorting}

According to the general theory of estimation (Eqs.\ (\ref{eq_a_2}) and
(\ref{eq_a_5})), given $\ve{z}_1$ and applying the optimum linear estimator
(feedforward processing), the correlation matrix of the error with respect to
the remaining, not yet known symbols $z_k$ is given as
\begin{eqnarray}						\label{eq_4_46}
\ve{\Phi}_{ee} &=& \left( \ve{C}_\ll^\H\ve{C}_\ll^{}
				\textstyle\frac{1}{\sigma_n^2}
		+ \ve{\Phi}_{22 \mid \ve{z}_1}^{-1} \right)^{-1}
\nonumber\\
&=& \sigma_n^2 \left( \ve{C}_\ll^\H\ve{C}_\ll^{} + \sigma_n^2( \ve{\Phi}_{22}^{}
		- \ve{\Phi}^{}_{21} \ve{\Phi}_{11}^{-1} \ve{\Phi}_{12}^{} )^{-1}
			\right)^{-1} \!\!\!\!\!
\nonumber\\
&=& \sigma_n^2 \left( \ve{C}_\ll^\H\ve{C}_\ll^{}
		+ \ve{A}_\ll^\H\ve{A}_\ll^{} \right)^{-1} \;.
\end{eqnarray}
The next symbol to be detected is the one, for which the corresponding main
diagonal element of $\ve{\Phi}_{ee}$ is minimum. Assuming the channel matrix
has been accordingly rearranged, the upper left main diagonal element is the
smallest and only the first row of the feedforward matrix is used to produce
a decision symbol. Otherwise, the respective rows are exchanged which is
kept track of in the permutation matrix $\ve{P}$.

\subsubsection{Feedback Matrix}

From (\ref{eq_a_1}) and using (\ref{eq_a_4}), (\ref{eq_a_5}), the feedback
filter follows immediately, too. The influence of the already detected symbols
has additionally to be canceled from the receive vector $\ve{y}$. This is done
by remodulating the vector $\ve{z}_1$ of decisions via $\ve{C}_\fl^{}$,
containing the first $l$ columns of $\ve{C}$. Moreover, the mean of $\ve{z}_2$
given $\ve{z}_1$ has to be taken into account (starting from
$\ve{\mu}_z = \ve{0}$). The task of the feedback filter is hence twofold:
to cancel the known interference and at the same time to predict the not yet
decided symbols from the known ones.

With the goal to have the cancellation point at the output of the feedforward
matrix, the feedback filter, when already $l$ symbols are known, calculates to
\begin{eqnarray}						\label{eq_4_50}
\ve{M}^{(l)} &=& \Big( \left( \ve{C}_\ll^\H \ve{C}_\ll^{}
			\textstyle\frac{1}{\sigma_n^2}
		+ \ve{\Phi}_{22 \mid \ve{z}_1}^{-1} \right)^{-1}
\\[-1mm]
	&& \qquad\qquad\qquad \cdot
		 \ve{C}_\ll^\H\ve{C}_\ll^{} \textstyle\frac{1}{\sigma_n^2}
			- \ve{I} \Big) \ve{\Phi}_{21}^{}\ve{\Phi}_{11}^{-1}
\nonumber\\
	&& \quad + \left( \ve{C}_\ll^\H \ve{C}_\ll^{}
		+ \sigma_n^2 \ve{\Phi}_{22 \mid \ve{z}_1}^{-1} \right)^{-1}
			\ve{C}_\ll^\H \ve{C}_\fl^{}
\nonumber\\
\intertext{and with the above abbreviations (partitioning of $\ve{\bar{C}}$),
	after straightforward manipulations, we arrive at}
&=& \Big( \left( \ve{C}_2^\H\ve{C}_2^{} + \ve{A}_2^\H\ve{A}_2^{} \right)^{-1}
	\ve{C}_2^\H\ve{C}_2 - \ve{I} \Big)
		\ve{\Phi}_{21}^{}\ve{\Phi}_{11}^{-1}
\nonumber\\[-2mm]
	&& \qquad\qquad
	+ \left( \ve{C}_2^\H\ve{C}_2^{} + \ve{A}_2^\H\ve{A}_2^{} \right)^{-1}
		\ve{C}_2^\H\ve{C}_1^{}
\nonumber\\
&=& \left( \ve{C}_2^\H\ve{C}_2^{} + \ve{A}_2^\H\ve{A}_2^{} \right)^{-1}
\nonumber\\[-2mm]
	&& \qquad\qquad \cdot \left( \ve{C}_2^\H\ve{C}_1^{} -
		\ve{A}_2^\H\ve{A}_2^{}\ve{\Phi}_{21}^{}\ve{\Phi}_{11}^{-1}
		\right) \;.
\end{eqnarray}
From (\ref{eq_4_42}), the correlation matrix is given as
\begin{eqnarray}						\label{eq_4_51}
\ve{\Phi}_{zz} &=& \left[ \matrix{ \ve{\Phi}_{11} & \ve{\Phi}_{12} \cr
			\ve{\Phi}_{21} & \ve{\Phi}_{22} } \right]
\nonumber\\
&=& \sigma_n^2 \left[ \matrix{
		\ve{A}_1^\H\ve{A}_1^{} & \ve{A}_1^\H\ve{A}_2^{} \cr
		\ve{A}_2^\H\ve{A}_1^{} & \ve{A}_2^\H\ve{A}_2^{} } \right]^{-1}
	\defeq \left[ \matrix{ \ve{O} & \ve{V} \cr
				  \ve{V}^\H & \ve{U} } \right]^{-1} \!\!\!\;.
\end{eqnarray}
Again using \cite[Eq.~(7.7.5)]{Horn:85}, we have
\begin{eqnarray}						\label{eq_4_52}
\ve{\Phi}_{11} &=& \left( \ve{O}
			- \ve{V}\ve{U}^{-1}\ve{V}^\H \right)^{-1}
\\
\ve{\Phi}_{21} &=& \left( \ve{V}^\H\ve{O}^{-1}\ve{V}
			- \ve{U} \right)^{-1} \ve{V}^\H\ve{O}^{-1}
\end{eqnarray}
and together with $\ve{A}_2^\H\ve{A}_2^{} = \ve{U}$, we arrive at
\begin{eqnarray}						\label{eq_4_53}
\ve{A}_2^\H\ve{A}_2^{}\ve{\Phi}_{21}^{}\ve{\Phi}_{11}^{-1}
&=& \ve{U} \left( \ve{V}^\H\ve{O}^{-1}\ve{V} - \ve{U} \right)^{-1}
\nonumber\\[-2mm]
&& \qquad\quad \cdot 
	\ve{V}^\H\ve{O}^{-1} \left( \ve{O} - \ve{V}\ve{U}^{-1}\ve{V}^\H \right)
\nonumber\\
&=& \ve{U} \left( \ve{V}^\H\ve{O}^{-1}\ve{V} - \ve{U} \right)^{-1}
\nonumber\\[-2mm]
&& \qquad\quad \cdot
	\left( \ve{U} - \ve{V}^\H\ve{O}^{-1}\ve{V} \right) \ve{U}^{-1}\ve{V}^\H
\nonumber\\
&=& -\ve{V}^\H \;.
\end{eqnarray}
In summary, the feedback matrix, when $l$ symbols are already known, is given
by
\begin{eqnarray}						\label{eq_4_54}
\ve{M}^{(l)}
	&=& \left( \ve{C}_2^\H \ve{C}_2^{} + \ve{A}_2^\H\ve{A}_2^{} \right)^{-1}
	\!\!\left( \ve{C}_2^\H\ve{C}_1^{} + \ve{A}_2^\H\ve{A}_1^{} \right) .
\end{eqnarray}
Assuming that the symbols $z_k$ are in the optimum ordering, as for the
feedforward matrix, since only a single next symbol (the currently best) is
decided, only the first row of the matrix $\ve{M}^{(l)}$ is actually used.
Note that the respective row of the feedback matrix $\ve{B}$ is obtained from
that row by appending a single one and then $\NT-l-1$ trailing zeros.

\subsection{Comparison and Discussion}

From the above derivations it is immediate that both perspectives on LRA MMSE
DFE lead to the same result. A comparison of (\ref{eq_4_10})---here deleting
the last $\NR$ columns to return from the augmented to the original channel
model---and (\ref{eq_4_45}) reveals that for both cases the feedforward
matrices are identical.

The sorting is based on (\ref{eq_4_12}) and (\ref{eq_4_46}), respectively.
As the feedforward processing is identical, this also holds for the error
variances or the norms of the filter vectors, proportional to these variances
and hence the same decision orders result.

Finally, the feedback filters are also identical; this is revealed by
comparing (\ref{eq_4_25}) and (\ref{eq_4_54}).

Hence, the straightforward application of the V-BLAST algorithm for sorted
QR decomposition to the extended channel matrix indeed results in the optimum
solution to LRA MMSE DFE. The ``trick'' behind this lies in the lower part
of the augmented matrix. Whereas for classical DFE the (scaled) identity matrix
is present, in case of LRA the inverse of the square root of the correlation
matrix of the vector $\ve{z}$ to be estimated is present (cf.\ (\ref{eq_3_11})
and (\ref{eq_4_42})). As shown, deleting columns and calculating the
feedforward matrix on this reduced channel matrix has the same effect as
updating the correlation matrix of the residual symbols.

The above derivation also reveals that in case of MMSE DFE for correlated
symbols the feedback matrix fulfills two tasks: the cancellation of the
interference of already detected symbols and some kind of prediction of the
still unknown information symbols from the symbols up to now known. In case
of white data symbols, only cancellation is required.

Numerical simulations reveal that it is more important to take the correlations
of the data symbols correctly into account than using a specific type of lattice
reduction. Conducting lattice reduction on the original channel matrix, and
using the resulting matrices $\ve{C}$ and $\ve{Z}$ to create an augmented
matrix on which the QR decomposition is done, performs only marginally worse
than starting rightaway with the augmented matrix. However, using the LLL
on the original channel requires less complexity and is independent of the
current SNR.

As the above derivation is valid for any channel model and any correlation of
the data, we can conclude that when performing MMSE DFE for correlated symbols,
optimum feedforward and feedback matrices and the optimum sorting can be
calculated via the V-BLAST algorithm. Thereby, the algorithm simply has to
work on an augmented channel matrix, which has the inverse of the square root
of the correlation matrix of the data symbols times the noise variance as its
lower part. In other words, all required matrices are obtained by performing a
sorted QR-type decomposition of this augmented channel matrix.
However, with regard to computational complexity this procedure is far from
optimum as the algorithm has to work on a matrix of approximately doubled
number of rows. Fortunately, the efficient ``fast V-BLAST algorithm'' proposed
in \cite[Table~II]{Benesty:03} can simply be modified to take correlated data
(correlation introduced via a matrix $\ve{Z}$) into account. Here, only the
computation of $\mathbf{R}$ and $\mathbf{Q}$ according to
\cite[Eqs.~(26) and (28)]{Benesty:03} has to be modified.
Using the initializations (notation from \cite{Benesty:03})
$\mathbf{R}_0 = \alpha\ve{Z}^\H\ve{Z}$ and
$\mathbf{Q}_0 = (1/\alpha)(\ve{Z}^\H\ve{Z})^{-1}$, this algorithm efficiently
delivers the same results as the ZF BLAST algorithm applied to the augmented
channel matrix.

LRA equalization for MIMO channels can be viewed as the counterpart to
partial-response signaling (PRS) \cite{Huber:92,Cioffi:95} for intersymbol
interference channels, see \cite{Fischer:06}. In both cases an integer
polynomial/matrix is split from the actual channel transfer function/matrix and
only the residual system is considered. Equalization is done towards the target
polynomial/unimodular matrix. The non-whiteness of the data sequence to
be detected has to be taken into account for MMSE equalization of PRS
(e.g.,\ \cite[Appendix~A]{Cioffi:95}); the same is true in LRA schemes.
However, in contrast to PRS, which is usually employed to achieve some desired
transmitter side characteristics (spectral zeros at DC or Nyquist frequency),
the use of LRA enables full diversity of the MIMO transmission system and hence
is the key to significantly improve error performance of uncoded transmission.

%
%
%

%%%%%%%%%%%%%%%%%%%%%%%%%%%%%%%%%%%%%%%%%%%%%%%%%%%%%%%%%%%%%%%%%%%%%%%%%%%%%%%
\section{Summary and Conclusions}
\label{sec5}

\noindent
Lattice-reduction-aided equalization optimized according to the MMSE criterion
of MIMO channels has been studied. For the first time it has been proven that
applying the zero-forcing BLAST algorithm to a suitably augmented channel
matrix---having the inverse of the square root of the correlation matrix of
the data symbols times the noise variance as its lower part---indeed results
in the optimum solution. It is more important to take the correlations of the
data correctly into account than what specific type of lattice reduction
actually is used.

Finally it should be noted that taking the \emph{uplink/down\-link du\-ali\-ty}
\cite{Tse:05} into account, instead of employing receiver-side equalization,
\emph{MMSE LRA precoding} \cite{Windpassinger:04trcom} can be performed.
The given results can immediately be transferred to this trans\-mitter-side
technique, which is of great importance in the multi-user downlink.

\appendix[Some Fundamentals of Estimation Theory]

\noindent
In this appendix, for convenience, two important properties on minimum
mean-squared error estimation of correlated and non-zero mean random variables
are reviewed from the literature.

First, we consider a vector $\ve{x}$ with (possibly) non-zero mean
$\ve{\mu}_x$ and covariance matrix $\ve{\Phi}_{xx}$. This vector is observed
through the matrix $\ve{H}$ and disturbed by (zero-mean) Gaussian noise
$\ve{n}$ with covariance matrix $\ve{\Phi}_{nn}$. Hence, the observation
$\ve{y} = \ve{H}\ve{x} + \ve{n}$ is present. The optimum linear estimator
for this setting is given by, e.g., \cite[Page~68]{Sayed:03}
\begin{eqnarray}						\label{eq_a_1}
\ve{\tilde{x}} &=& \left( \ve{H}^\H \ve{\Phi}_{nn}^{-1} \ve{H}^{}
		+ \ve{\Phi}_{xx}^{-1} \right)^{-1}\!\!\ve{H}^\H\ve{\Phi}_{nn}^{-1}
		(\ve{y} - \ve{H}\ve{\mu}_x) + \ve{\mu}_x
\nonumber\\
&=& \left( \ve{H}^\H \ve{\Phi}_{nn}^{-1} \ve{H}^{}
		+ \ve{\Phi}_{xx}^{-1} \right)^{-1} \!\!
			\ve{H}^\H\ve{\Phi}_{nn}^{-1}\ve{y}
\\
&& \mbox{} - \left( \left( \ve{H}^\H \ve{\Phi}_{nn}^{-1} \ve{H}^{}
		+ \ve{\Phi}_{xx}^{-1} \right)^{-1} \!\!
			\ve{H}^\H\ve{\Phi}_{nn}^{-1}\ve{H}
				- \ve{I}\right) \ve{\mu}_x \;.
\nonumber
\end{eqnarray}
The covariance matrix of the resulting estimation error can be written as
\begin{equation}						\label{eq_a_2}
\ve{\Phi}_{ee} = \left( \ve{\Phi}_{xx}^{-1} + \ve{H}^\H \ve{\Phi}_{nn}^{-1}
		\ve{H}^{} \right)^{-1} \;.
\end{equation}

Second, assume a multivariate Gaussian distribution (random vector $\ve{w}$)
of dimension $Q$ with mean $\ve{\mu}_w$ and covariance matrix $\ve{\Phi}_{ww}$.
Let the random vector, the mean vector, and the covariance matrix be
partitioned according to
\begin{equation}						 \label{eq_a_3}
\ve{w} = \left[ \matrix{ \ve{w}_1 \cr \ve{w}_2 } \right]
	,\;
\ve{\mu}_w = \left[ \matrix{ \ve{\mu}_1 \cr \ve{\mu}_2 } \right]
	,\;
\ve{\Phi}_{ww} = \left[ \matrix{ \ve{\Phi}_{11} & \ve{\Phi}_{12} \cr
                            \ve{\Phi}_{21} & \ve{\Phi}_{22} } \right]
\end{equation}
where the dimensions of the upper and left parts are $q$, e.g.,
$\dim(\ve{w}_1) = \dim(\ve{\mu}_1) = q$, $\dim(\ve{\Phi}_{11}) = q\times q$,
etc.

Having already knowledge on the first $q$ components of the random vector
$\ve{w}$---i.e., the vector $\ve{w}_1$---the mean and the covariance matrix
for the residual $Q-q$ variables (vector $\ve{w}_2$), conditioned on the
knowledge $\ve{w}_1$, calculate to
\begin{eqnarray}						 \label{eq_a_4}
\ve{\mu}_{2 \mid \ve{w}_1^{}} &=& \ve{\mu}_2^{} +
	\ve{\Phi}_{21}^{} \ve{\Phi}_{11}^{-1} \left(\ve{w}_1 - \ve{\mu}_1\right)
\\								 \label{eq_a_5}
\ve{\Phi}_{22 \mid \ve{w}_1}^{} &=& \ve{\Phi}_{22}^{}
	- \ve{\Phi}^{}_{21} \ve{\Phi}_{11}^{-1} \ve{\Phi}_{12}^{} \;.
\end{eqnarray}
Note that the new covariance matrix is the \emph{Schur complement} of
$\ve{\Phi}_{11}$ in $\ve{\Phi}$; it does not depend on the actual value of
$\ve{w}_1$. Note additionally that both quantities can be obtained
in one step or successively in $q$ steps, each time assuming additional
knowledge of a single symbol.

\vspace*{-1mm}
%%%%%%%%%%%%%%%%%%%%%%%%%%%%%%%%%%%%%%%%%%%%%%%%%%%%%%%%%%%%%%%%%%%%%%%%%%%%%%%

%
%
%

%%%%%%%%%%%%%%%%%%%%%%%%%%%%%%%%%%%%%%%%%%%%%%%%%%%%%%%%%%%%%%%%%%%%%%%%%%%%%%%

\begin{thebibliography}{00}

\bibitem{Benesty:03}
J. Benesty, Y. Huang, J. Chen.
\newblock A Fast Recursive Algorithm for Optimum Sequential Signal Detection
        in a BLAST System.
\newblock \emph{IEEE Transactions on Signal Processing}, Vol.~51, No.~7,
        pp.~1722--1730, July 2003.

\bibitem{Cioffi:95}
J.M. Cioffi, G.P. Dudevoir, M.V. Eyubo\v{g}lu, and G.D. Forney.
\newblock MMSE Decision-Feedback Equalizers and Coding---Part I:
	Equalization Results, Part II: Coding Results.
\newblock \emph{IEEE Transactions on Communications}, Vol.~43, No.~10,
	pp.~2582--2604, Oct.\ 1995.

\bibitem{Fischer:02}
R.F.H. Fischer.
\newblock \emph{Precoding and Signal Shaping for Digital Transmission},
\newblock John Wiley \& Sons, New York, 2002.

\bibitem{Fischer:03reell}
R.F.H. Fischer, C. Windpassinger.
\newblock Real- vs.\ Complex-Valued Equalisation in V-BLAST Systems.
\newblock \emph{Electronics Letters}, Vol.~39, No.~5, pp.~470--471, Mar.\ 2003.

\bibitem{Fischer:06}
R.F.H. Fischer, C. Siegl.
\newblock On the Relation between Lattice-Reduction-Aided Equalization and
	Partial-Response Signaling.
\newblock \emph{International Zurich Seminar (IZS)}, pp.~34--37,
	Zurich, Switzerland, Feb.\ 2006.

\bibitem{Gan:09}
Y.H. Gan, C. Ling. W.H. Mow.
\newblock   Complex Lattice Reduction Algorithm for Low-Complexity
        Full-Diversity MIMO Detection.
\newblock \emph{IEEE Transactions on Signal Processing}, Vol.~57, No~7,
        pp.~2701--2710, July 2009.

\bibitem{Ginis:02}
G. Ginis, J.M. Cioffi.
\newblock On the relation between V-BLAST and the GDFE.
\newblock IEEE Communications Letters, Vol.~5, No.~9, pp.~364--366,
	Sept.\ 2001.

\bibitem{Golden:99}
G.D. Golden, G.J. Foschini, R.A. Valenzuela, P.W. Wolniansky.
\newblock Detection Algorithm and Initial Laboratory Results Using V-BLAST
        Space-Time Communication Architecture.
\newblock \emph{Electronics Letters}, Vol.~35, No.~1, pp.~14--15, Jan.\ 1999.

\bibitem{Hassibi:00}
B. Hassibi.
\newblock An Efficient Square-Root Algorithm for BLAST.
\newblock \emph{IEEE International Conference on Acoustics, Speech, and
        Signal Processing} pp.~737--740, Istanbul, Turkey, June 2000.

\bibitem{Horn:85}
R.A. Horn, C.R. Johnson.
\newblock \emph{Matrix Analysis}.
\newblock Cambridge University Press, Cambridge, UK, 1985.

\bibitem{Huber:92}
J.~Huber.
\newblock \emph{Trelliscodierung}.
\newblock Springer Verlag, Berlin, Heidelberg, 1992.
\newblock (in German).

\bibitem{Lenstra:82}
A.K. Lenstra, H.W. Lenstra, L. Lov{\'a}sz.
\newblock Factoring polynomials with rational coefficients,
\newblock \emph{Mathematische Annalen}, Vol.~261, No.~4, pp.~515--534, 1982.

\bibitem{Murugan:06}
A.D. Murugan, H. El Gamal, M.O. Damen, G. Caire.
\newblock A Unified Framework for Tree Search Decoding: Rediscovering the
	Sequential Decoder.
\newblock \emph{IEEE Transactions on Information Theory}, Vol.~53, No.~3,
	pp.~933--953, Mar.\ 2006.

\bibitem{Sayed:03}
A.H. Sayed.
\newblock \emph{Fundamentals of Adaptive Filtering},
\newblock John Wiley \& Sons, New York, 2003.

\bibitem{Taherzadeh:07}
M. Taherzadeh, A. Mobasher, A.K. Khandani.
\newblock LLL Reduction Achieves the Receive Diversity in MIMO Decoding.
\newblock \emph{IEEE Transactions on Information Theory}, Vol.~53, No.~12,
        pp.~4801--4805, Dec.\ 2007.

\bibitem{Tse:05}
D. Tse, P. Viswanath.
\newblock \emph{Fundamentals of Wireless Communication}.
\newblock Cambridge Universty Press, Cambridge, UK, 2005.

\bibitem{Windpassinger:03}
C. Windpassinger, R.F.H. Fischer.
\newblock Low-Complexity Near-Maxi\-mum-Likeli\-hood Detection and
        Precoding for MIMO Systems using Lattice Reduction.
\newblock \emph{IEEE Information Theory Workshop}, pp.~345-348,
        Paris, France, Mar./Apr.\ 2003.

\bibitem{Windpassinger:04}
C. Windpassinger.
\newblock \emph{Detection and Precoding for Multiple Input
        Multiple Output Channels}.
\newblock Dissertation, Erlangen, June 2004.

\bibitem{Windpassinger:04trcom}
C. Windpassinger, R.F.H. Fischer, J.B. Huber.
\newblock {Lattice-Re\-duc\-tion-Aided Broadcast Precoding}.
\newblock \emph{IEEE Transactions on Communications}, Vol.~52, No.~12,
        pp.~2057--2060, Dec.\ 2004.

\bibitem{Wuebben:04}
D. W{\"u}bben, R. B{\"o}hnke, V. K{\"u}hn, K.D. Kammeyer.
\newblock {Near-Maxi\-mum-Likeli\-hood Detection of MIMO Systems using
        MMSE-Based Lattice Reduction}.
\newblock \emph{IEEE International Conference on Communications},
        pp.~798--802, Paris, France, June 2004.

\bibitem{Yao:02}
H. Yao, G.W. Wornell.
\newblock Lattice-Reduction-Aided Detectors for MIMO Communication Systems.
\newblock \emph{IEEE Global Communications Conference}, Taipei, Taiwan,
        Nov.\ 2002.

\end{thebibliography}
\end{document}